# Enhanced Operational Stability of Perovskite Light-Emitting Electrochemical Cells Leveraging Ionic Additives

*Aditya Mishra, Masoud Alahbakhshi, Ross Haroldson, Lyndon D. Bastatas, Qing Gu, Anvar A. Zakhidov and Jason D. Slinker\**

A. Mishra, Prof. J. D. Slinker
Department of Materials Science and Engineering, The University of Texas at Dallas, 800 West Campbell Road, Richardson, Texas 75080-3021, United States.
E-mail: slinker@utdallas.edu

M. Alahbakhshi, Prof. Q. Gu
Department of Electrical and Computer Engineering, The University of Texas at Dallas, 800 West Campbell Road, Richardson, Texas 75080-3021, United States.

R. Haroldson, Prof. A. A. Zakhidov, Prof. J. D. Slinker
Department of Physics, The University of Texas at Dallas, 800 West Campbell Road, Richardson, Texas 75080-3021, United States.

Prof. Lyndon D. Bastatas
Physics Department, Western Mindanao State University, Normal Road, Baliwasan, Zamboanga City 7000 Philippines

Prof. A. A. Zakhidov
NanoTech Institute, The University of Texas at Dallas, 800 West Campbell Road, Richardson, Texas 75080-3021, United States.

Prof. A. A. Zakhidov
Department of Nanophotonics and Metamaterials, ITMO University, St. Petersburg, Moscow, Russia.

Prof. A. A. Zakhidov
Laboratory of Advanced Solar Energy, NUST MISiS, Moscow, 119049, Russia

Keywords: electroluminescence, operational lifetime, degradation, electrochemical impedance spectroscopy, mixed conductors

Hybrid perovskites are emerging as highly efficient materials for optoelectronic applications, however, the operational lifetime has remained a limiting factor for the continued progress of perovskite light emitting devices such as light emitting diodes (LEDs) and perovskite light emitting electrochemical cells (PeLECs). In this work, PeLECs utilizing an optimized fraction of $LiPF_6$ salt additive exhibit enhanced stability. At 0.5 wt% $LiPF_6$, devices exhibit 100 h operation at high brightness in excess of 800 cd m$^{-2}$ under constant current driving, achieving a maximum luminance of 3260 cd m$^{-2}$ and power efficiency of 9.1 Lm W$^{-1}$. This performance extrapolates to a 6700 h luminance half-life from 100 cd m$^{-2}$, a 5.6-fold improvement over devices with no lithium salt additive. Analysis under constant voltage driving reveals three current regimes, with lithium addition strongly enhancing current in the second and third regimes. The third regime correlates degradation of luminance with decreased current. These losses are mitigated by LiPF$_6$ addition, an effect postulated to arise from preservation of perovskite structure. To further understand lithium salt addition, electrochemical impedance spectroscopy with equivalent circuit modeling is performed. Electrical double layer widths from ionic redistribution are minimized at 0.5 wt% LiPF$_6$ and inversely correlate with efficient performance.

**1. Introduction**

Solution processed organic-inorganic metal halide perovskites with the chemical form of ABX$_3$ have emerged as next generation light emitting materials.[1] Metal halide perovskites exhibit unique optoelectronic characteristics such as high photoluminescence quantum efficiency (PLQE), efficient electroluminescence (EL) with narrow full width half-maximum (FWHM), ease of band-gap tunability, photon recycling and low temperature processability.[2] In recent years, perovskite light emitting diodes (PeLEDs) have improved in luminance from 10 cd/m$^2$ to ≈ 10$^4$ cd/m$^2$ with various external quantum efficiencies (EQE) from 0.1 to ≈ 10%.[3] Despite these promising results and considerable research on PeLEDs, operational stability is still a roadblock for their insertion into most practical applications.[1-2, 4] The majority of these PeLEDs utilize a complex multilayer device architecture prepared with various passivation and crystallization techniques to minimize non-radiative pathways in the emissive layer.[2, 5] The lifetime of multilayer PeLEDs has been reported in the range of 0.1 to 100 h (See Table S1).[3d, 4a, 6] This limited lifetime of PeLEDs has been explained in the context of non-uniform morphology of the emissive layer, electrode delamination, and high densities of interfacial defect traps.[4a, 4b, 7]

Recently, ion migration has been demonstrated to significantly influence the performance of metal halide perovskite devices.[8] High capacitance at low frequencies, electrode polarization, scan rate dependent hysteresis, and EL color changes during operation are all attributed to ion motion.[9] Rationale for ion motion includes phenomena such as a low activation energy of diffusion for halide ions and charged point defects, stochiometric error induced extrinsic defects, and polycrystalline soft lattices with high densities of grain boundaries.[9c, 10] These studies collectively establish metal halide perovskites as mixed conductors with both electronic and ionic conductivities. The interplay between electronic and ionic currents provides a new pathway to utilize metal halide perovskites as emissive materials in light emitting electrochemical cells (PeLEC).

Recent efforts have aimed to realize high performance PeLEC devices utilizing the benefits of ionic redistribution. Aygüler et al. blended formamidium lead bromide nanoparticles with a trimethylolpropane ethoxylate polyelectrolyte and lithium triflate salt, achieving luminance on the order of 1 cd m$^{-2}$.[11] Li et al. characterized $CH_3NH_3PbBr_3$:PEO single layer devices and achieved up to 4000 cd m$^{-2}$ max luminance, with device performance indicative of both PeLEDs and polymer LECs.[6f] Mixed conductivity was further explored by Zhang et al., who suggested that perovskite functioned as a solid electrolyte and demonstrated light emission in PeLEC with both forward and reverse biases.[12] Additionally, they observed halide migration followed by ionic accumulation, which resulted in large interfacial capacitance in the device. Furthermore, Puscher et al. studied the migration and rearrangement dynamics of different ionic species in perovskite nano-particle based thin film devices and presented evidence for electric double layer (EDL) formation followed by charge injection.[13] Andričević et al. created single crystal PeLECs with vertically aligned carbon nanotube electrodes and obtained fluctuating emission with a maximum luminance of 1800 cd m$^{-2}$.[14] Low temperature operation revealed voltage-dependent EL spectral shifts consistent with effects of ionic redistribution.

Recently, our group demonstrated bright and effectual PeLECs with a simple single layer architecture, leveraging a lithium ionic additive (LiPF$_6$) to achieve high luminance of ~15000 cd m$^{-2}$.[15] The device has a simple device architecture consisting of ITO/CsPbBr$_3$:PEO:LiPF$_6$/InGa eutectic. Advantages of LiPF$_6$ salt in the CsPbBr$_3$:PEO active layer include uniform and pinhole free morphology, high PLQE, stable PL dynamics, EL stability, low hysteresis and high efficiency. In pristine metal halide perovskites (i.e. without any additives), the disparity between the diffusion activation energies of halide ions and A$^+$ cations causes an imbalance in space charge density near the electrodes[13, 15] and consequently results in unequal charge carrier (e$^-$ and h$^+$) concentrations in the active layer (See Table S2).[16] To resolve this imbalance, external lithium ions were added as mobile cations capable of enhancing cation redistribution, facilitating better n-type doping and EDL formation at the cathode.[17] Further study is needed to clarify the details of EDL formation in response to lithium addition in efficient PeLECs.

In the present study, we utilize Li additives and constant current driving to achieve PeLECs with the highest operational stability reported to date. Devices are also driven with constant voltage to understand the relationships between current, light emission, and degradation. Electrical double layer formation and other key parameters critical for PeLEC device operation were characterized by electrochemical impedance spectroscopy. In addition, equivalent circuit modelling of PeLEC provided a new outlook of ionic migration in perovskite light emitting devices.

## 2. Results and Discussion
### 2.1. Mechanism of PeLEC

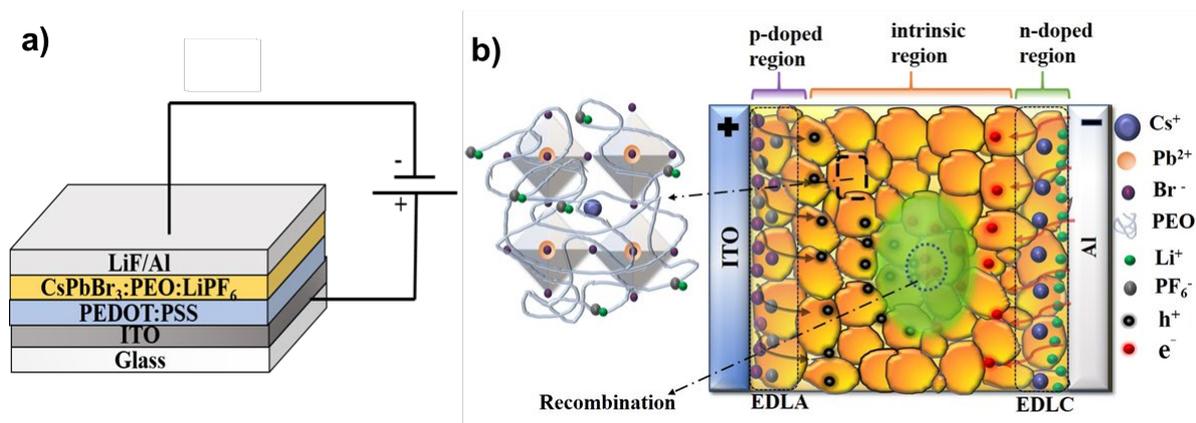

**Figure 1.** a) Illustration of PeLEC device architecture (ITO/PEDOT:PSS/CsPbBr$_3$: PEO:x%LiPF$_6$/LiF-Al). (b) Steady-state PeLEC operation mechanism with ions accumulated at the electrodes and light emission upon current injection, where EDLA and EDLC are electric double layer at anode (ITO) and cathode (Al) respectively.

We fabricated PeLECs as shown in Figure 1a. We prepared CsPbBr$_3$:PEO:LiPF$_6$ (1:0.8:x% w/w) active layer LECs with an indium tin oxide (ITO) anode and an aluminum (Al) cathode along with poly(ethylenedioxythiophene) poly(styrene sulfonate) PEDOT:PSS, which serves to smooth the ITO surface for higher device reliability. Figure 1b illustrates the PeLEC operational mechanism under an applied bias. LEC device operation has been explained in the following distinct stages: 1) Under an external electric field, anions (Br$^-$ and PF$_6^-$) and cations (Li$^+$ and Cs$^+$) drift towards anode and cathode respectively and accumulate near each electrode. 2) The accumulation of these ions at the electrodes leads to the formation of EDLs and subsequent p and n-type doping. 3) The potential barrier width at each electrode-active layer interface is reduced due to EDL formation, facilitating charge injection and improved transport through p and n-doped regions of an in-situ formed p-i-n junction. 4) Injected charges are transported through the bulk of the perovskite film, followed by radiative recombination in the intrinsic region of the p-i-n structure.
.

## 2.2. PL and EL from Perovskite Thin Films and LECs

To characterize the optical bandgap of pristine CsPbBr$_3$:PEO spin coated thin films, we performed PL and absorption spectroscopy as shown in Figure 2a. We observed a PL peak and an absorption edge located at 520 nm, corresponding to a bandgap of ~2.3 eV. This demonstrates that the film does not undergo a band gap shift from spin coating.

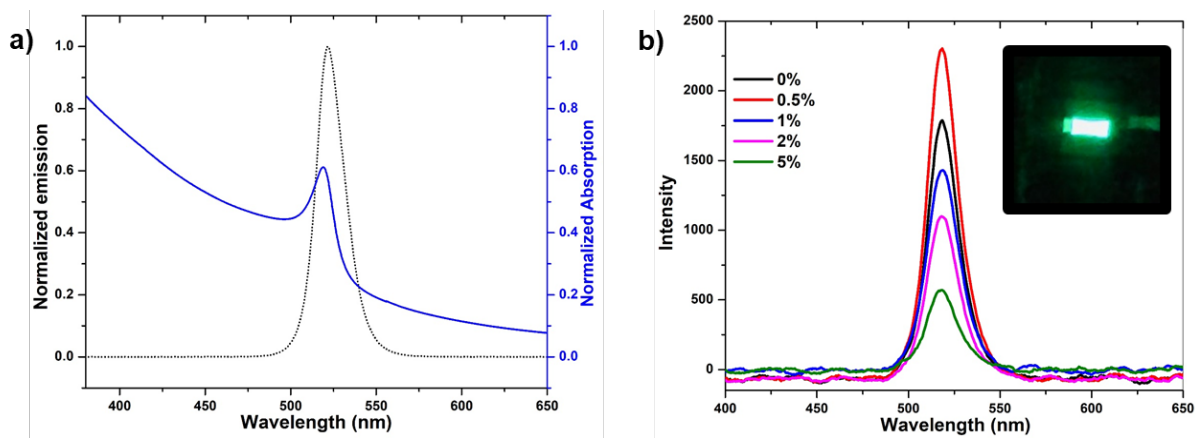

**Figure 2.** a) Normalized photoluminescence (black) at an excitation wavelength of 400 nm and absorption spectra (blue). b) Electroluminescence spectra of ITO/PEDOT:PSS /CsPbBr$_3$:PEO:LiPF$_6$/LiF/Al with various concentration (w/w) of LiPF$_6$. The inset shows the electroluminescence of a 0.5% LiPF$_6$ at 3.5 V.

In order to understand the influence of the Li salt on device emission color, we measured the EL spectra of fabricated devices (ITO/PEDOT:PSS/CsPbBr$_3$:PEO:x% LiPF$_6$/LiF/Al) with various LiPF$_6$ concentrations at 3.5 V, as shown in Figure 2b. All devices exhibited EL peaks at 520 nm, identical to the PL peak wavelength in Figure 2a. Furthermore, the EL intensity was found to be highest for films with 0.5 wt% LiPF$_6$. This enhancement in EL agrees with our pervious observation of enhanced EL with an optimized LiPF$_6$ concentration under constant voltage operation.[15]

## 2.3 PeLEC Device Performance with LiPF$_6$ Additives Under Constant Current

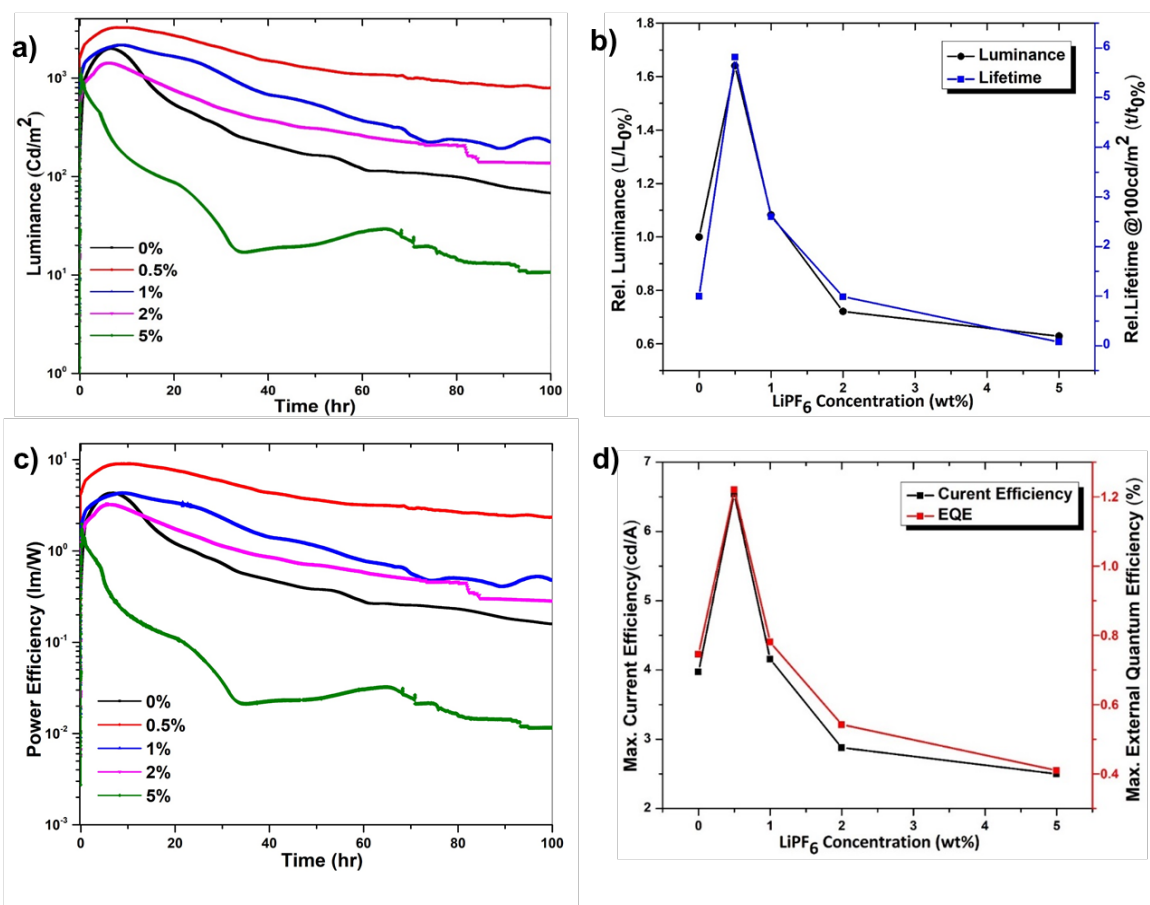

**Figure 3.** Luminance, efficiency, and lifetime analysis of PeLEC devices with various amounts of LiPF$_6$ under constant current driving (0.050 A/cm$^2$). a) Luminance vs time for PeLECs with various LiPF$_6$ concentrations under constant current driving. b) Relative maximum luminance and relative lifetime (half-life) extrapolated at 100 cd/m$^2$ vs LiPF$_6$ concentration for PeLEC devices, each normalized against a pristine PeLEC with 0% LiPF$_6$. c) Power efficiency vs time for PeLECs with various LiPF$_6$ concentrations. d) Maximum current efficiency and maximum power efficiency vs LiPF$_6$ concentration for PeLEC devices under constant current driving.

To characterize the device lifetime, we performed contstant current study of PeLECs with different LiPF$_6$ concentrations. Constant current operation leads to favorably lower operational voltages as ionic redistribution and EDL formation facilitates facile charge injection (See Figure S1). In LECs under operation, ions redistribute, and charge injection is improved exponentially by this effect. These actions decrease the resistance of the device due to the increased density of charge carriers. For constant voltage driving, the applied bias remains the same in spite of the decreased resistance, which can lead to electrochemical breakdown. For constant current driving, the voltage lowers as the effective device resistance is lowered for greater electrochemical stability.

Figure 3a presents the luminance vs time of PeLECs with different concentrations of LiPF$_6$ under a constant driving current of 0.05 A cm$^{-2}$. In all cases, the PeLEC rapidly emits significant EL until luminance maxima is reached, followed by a decreased luminance loosely following biexponential decay with time. The device with 0.5 wt% LiPF$_6$ achieved the highest maximum luminance of 3260 cd m$^{-2}$, significantly surpassing the 1980 cd m$^{-2}$ figure of the pristine 0 wt% LiPF$_6$ device. The luminance maxima vs LiPF$_6$ concentration shown in Figure 3b exhibits a sharp peak at the optimal 0.5 wt%. Close inspection of the luminance vs time curves in Figure 3a also reveals that the greatest operational stabilty is achieved with the 0.5 wt% LiPF$_6$ device. The decay portion can be fit with two exponential factors of the form A$e^{(-t/\tau)}$, with the initial decay region (10-40 h) following $\tau_1$ = 54 h and the latter region (40-100 h) fit by $\tau_2$ = 190 h (See supporting information Figure S2). By comparison, exponential fits of the 0 wt% LiPF$_6$ device showed $\tau_1$ = 8 h and $\tau_2$ = 23 h, both considerably lower. Overall, this contributes to 100 h of continuous operation in excess of 800 cd m$^{-2}$ for the 0.5 wt% LiPF$_6$ device.

It is challenging to directly compare the device lifetime at various LiPF$_6$ concentrations, given that they operate at different luminance levels. Alternatively, we present the extrapolated half-lives (time to decay to half of the maximum luminance) at a common 100 cd m$^{-2}$ initial luminance using the equation:

$$T_2 = T_1 \left(\frac{L_1}{L_2}\right)^{A_F} \tag{1}$$

where $T_2$ is the extrapolated half-life from an initial luminance of $L_2$ = 100 cd m$^{-2}$, $T_1$ is the experimentally measured half-life at the experimentally measured maximum luminance of $L_1$, and $A_F$ is a dimensionless exponential acceleration factor taken to be 1.5-1.7 from prior experimental observations.[3d, 18] To compare all devices, we extracted the extrapolated luminance half-life using a modest $A_F$ of 1.5, with results normalized to the pristine device in Figure 3b. The PeLEC device with 0.5% LiPF$_6$ showed the longest extrapolated lifetime at 100 cd m$^{-2}$ of 6700 h, 5.8 times longer than the pristine device and a highly competitive factor among published PeLED and PeLEC works (See Table S1).

In Figure 3c the power efficiency as a function of time also showed similarly improved values, peaking at 9.1 Lm W$^{-1}$ for 0.5% LiPF$_6$, a twofold improvement over the pristine device. Improved stabilty is also evident for the 0.5% LiPF$_6$ power efficiency curve, facilitated by constant current operation. Improvement in current efficiency (CE) and EQE followed the same trend as luminance and lifetime as shown in Figure 3d. A maximum CE of 6.6 cd A$^{-1}$ and EQE of 1.2% were achieved at 0.5% LiPF$_6$ concentration, each 60-70% higher than the device without LiPF$_6$ (CE = 3.9 cd A$^{-1}$, and EQE = 0.74%).

## 2.4 PeLEC Device Performance with LiPF$_6$ Additives Under Constant Voltage

Additionally, we studied luminance vs time at a constant voltage (3.5V) as shown in Figure S3, where we observed maximal performance (2350 cd m$^{-2}$, 8.4 cd A$^{-1}$, 1.7% EQE) from the device with the optimized concentration of LiPF$_6$ (0.5%). Interestingly, the higher current density observed from the PeLEC with 0.5% LiPF$_6$ indicates improvement in the conductivity of the film. Faster ionic response was also observed in the cyclic J-V sweep, where the device with 0.5% LiPF$_6$ experienced lower hysteresis than the pristine device.

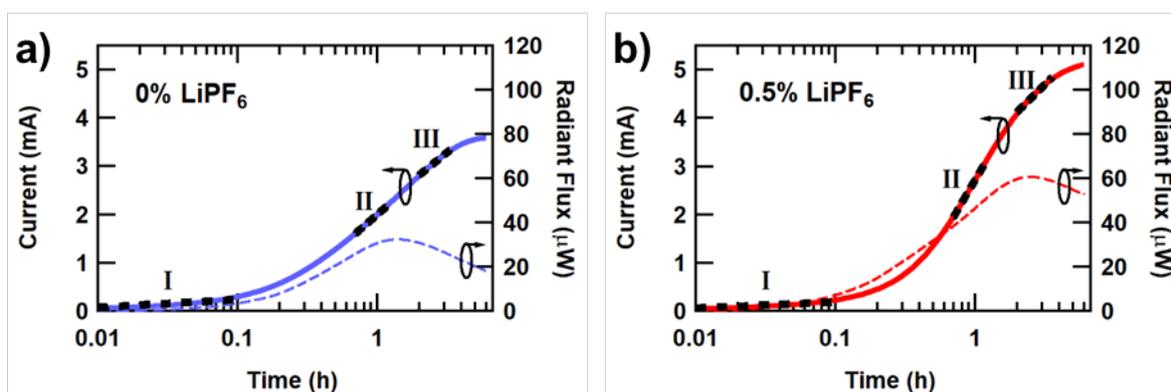

Figure 4. a) Current (solid blue lines) and radiant flux (dashed blue lines) vs time at constant 3.5 V for a pristine PeLEC. Dashed black lines are fits to various regions of interest. b) Current

(solid red lines) and radiant flux (dashed red lines) vs time at constant 3.5 V for a PeLEC with 0.5 wt% LiPF$_6$. Dashed black lines are fits to various regions of interest.

We also analyzed the dynamics of EDL formation near the electrode through current vs time analysis at constant voltage driving as shown in Figures 4a and 4b. To further understand the nature of the processes, we also recorded and plotted the radiant flux vs time for these measurements. We applied a constant 3.5 V to PeLEC devices with 0 wt% and 0.5 wt% of LiPF$_6$ and observed a LEC feature where three regimes in the slope of current as a function of time are evident.[13, 15] In particular, these three regimes can be fit by linear equations of the form:

$$i(t) = m * \log t + b, \tag{5}$$

where *m* and *b* are fitting constants associated with the slope and y axis intercept of each line, respectively. To understand the role of the ionic additive in the dynamics of EDL formation, we considered the fundamental processes occurring at these three characteristic timescales for pristine and salt-enhanced devices. For pristine devices (Figure 5a), the earliest time region (t < 0.2 h) can be attributed to the accumulation of Br$^-$ anions at anode (ITO) and anodic EDL formation, as halide ionic defects exhibit the lowest activation energy for ion migration (See Table S2).[10b, 13] A shallow current increase (slope $m_I$ = 0.16) is observed in this range, and little change is seen in the light emission, confirming that one carrier type is dominating in this region. The second time regime (0.7 h < t < 1.2 h) corresponds to EDL formation at the cathode by cations, facilitated by overcoming the activation energy of cesium cationic defects.[10b, 13] Indeed, the bipolar charge injection in this regime is confirmed by the large increase in radiant flux and a sharp current increase ($m_{II}$ = 2.55). This also indicates that significant p-i-n junction/recombination zone formation is occurring in the second regime. In the third regime, (2h < t < 3.2h) the current slope ($m_{III}$ = 2.35) is smaller than the second regime. Previously, this regime has been primarily attributed to p-i-n junction formation,[13, 15] but in considering the radiant flux curve, it is clear that this range correlates with a *decreasing* rate of radiant flux.

Instead, formation of the p-i-n junction or recombination zone must largely occur in the second regime, when radiant flux dramatically increases, and is completed at the onset of the third region. Alternatively, a degrading process that decreases both the rate of current flow and the rate of radiant flux emission distinguishes the third regime.

In considering the PeLEC with 0.5 wt% LiPF$_6$ (Figure 4b), all three regimes identified in the pristine device are also observed. The current increase in the initial region corresponding to anodic EDL formation is comparable ($m_I \approx 0.13$) to the pristine device. However, there is a significant increase in the slope in the second regime ($m_{II} \approx 4.97$) and a higher radiant flux. This is the regime corresponding to cathodic EDL formation, precisely the electrode benefitted by Li$^+$ ions. Li addition strongly enhances electron injection through compact and balanced double layer formation (through the smaller Li$^+$ size compared to Cs$^+$), as revealed by EIS analysis discussed in the next section, benefitting light emission and efficiency metrics. LiPF$_6$ addition also improves slope in the third region ($m_{III} \approx 2.86$) over the pristine device, indicating that the degradation impact on conductivity is mitigated.

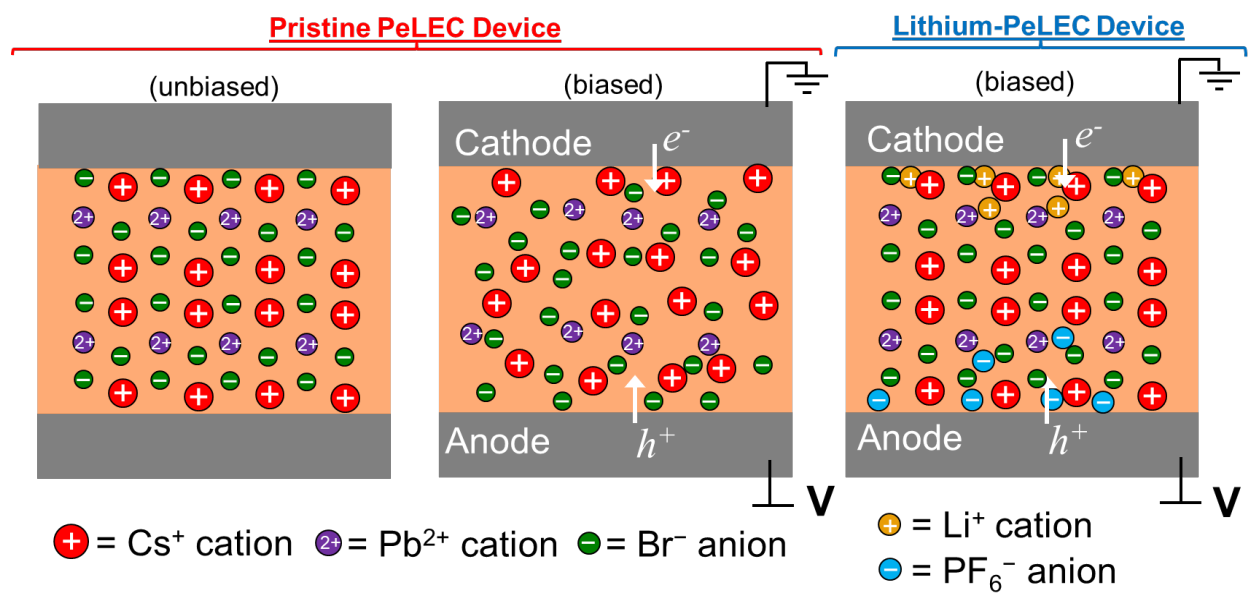

Figure 5. Conceptual drawing of a PeLEC before and after bias for a pristine device, as well as a lithium salt enhanced PeLEC. In the unbiased pristine device, the ions of the perovskite film are ordered in crystal domains. Upon application of a bias, these ions redistribute, disrupting the inherent structure. Alternatively, for the PeLEC device with lithium salt, the added ions redistribute and help preserve the order of the perovskite.

At this point, we pose the question: why would ion addition bring about better stability in a PeLEC? As illustrated in Figure 1, $CsPbBr_3$ perovskite films are polycrystalline, and crystallinity contributes to high conductivity in support of efficient device operation. We consider the effects of the applied bias on the ionic redistribution and the crystal structure of the perovskite in a pristine film PeLEC with the illustration shown in Figure 5. Before a bias is applied, the material is ordered, with a regular arrangement of $Cs^+$ and $Pb^{2+}$ cations and $Br^-$ anions. Once a bias is applied, ions of the perovskite redistribute, with positive ions attracted to the cathode and negative ions drifting toward the anode. This action assists charge injection due to the accumulation of ionic space charge near the electrodes. However, this ionic redistribution can negatively impact conductivity due to distortion of the crystal structure of perovskite, and likewise lower light emission due to the formation of vacancies and other traps. This ion redistribution leads to hysteresis, phase separation, and, in extreme cases, device failure possibly due to the formation of $PbBr_2$ insulating layers from accumulated ions at interfaces.[8b, 19]

The situation is different with added salts such as $LiPF_6$. In this case, the added mobile $Li^+$ and $PF_6^-$ ions redistribute more favorably than the intrinsic ionic species and largely preserve the inherent structure of the perovskite film. The redistribution of these small, mobile ions builds up high EDLs and electric fields at the electrodes, leading to a balanced charge injection. High conductivity and efficient emission are retained due to the preservation of the underlying perovskite structure. This also leads to devices of higher stability as shown in Figure 3, as well as greater reversibility as shown in Figure S4. In this manner, the added ions in the PeLEC structure balance the necessary charge redistribution in response to the applied field and limit the impact of the applied field on the perovskite structure.

Our previous work provides support of this view that lithium preserves inherent perovskite structure.[15] X ray photoelectron spectroscopy (XPS) study showed that Li salt addition increases the bonding energies of the inherent perovskite ions. X ray diffraction (XRD) demonstrated that the Li salt reduces the orthorhombic *Pnma* crystallite size by 20-25%, consistent with stronger bonding. Scanning electron microscopy (SEM) showed that the optimal 0.5 wt% LiPF$_6$ perovskite film produced a maximal grain size. Furthermore, prior electric force microscopy demonstrated that LiPF$_6$ addition reduced the electric field in the bulk of the LEC relative to a pristine device, [17c] an action that would suppress further bulk ion motion. All of these effects are consistent with a model of Li$^+$ and PF$_6^-$ ion motion that serves to preserve the inherent crystal structure of the perovskite.

**2.5 Electrochemical Impedance Spectroscopy (EIS)**

To extend our understanding of the ionic migration and EDL formation in PeLECs, we performed EIS at 0 V DC bias (shorted electrodes). The difference in cathode and anode workfunctions causes a built-in electric field within the device. At 0V, the ions redistribute to cancel out the built-in electric field, facilitating the EDL formation at the interfaces. By applying a small AC voltage perturbation, we can probe this ionic distribution effectively without operating the device. By minimizing the electronic carrier injection, we distinguish the ionic transport from the electronic carrier transport.

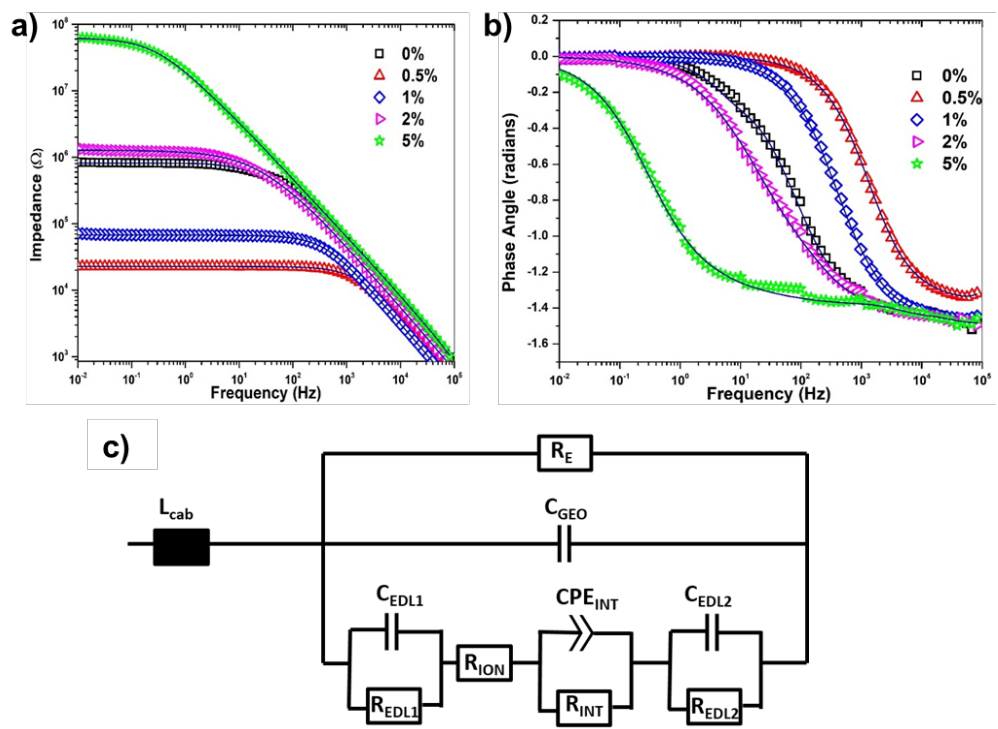

Figure 6. EIS of PeLEC devices with various fractions of lithium salt additives. (a) Impedance vs frequency data for PeLEC devices with various concentrations of $LiPF_6$. Solid lines are fits to the data based on the equivalent circuit. (b) Phase vs frequency data for PeLEC devices with various concentrations of $LiPF_6$. Solid lines are fits to the data based on the equivalent circuit. (c) Equivalent circuit used for EIS fitting. $L_{cab}$ is the inductance of the external cables; $R_E$ is the total electrical resistance of the active layer; $C_{GEO}$ is the geometric capacitance; $C_{EDL1}$ and $C_{EDL2}$ are capacitors representing the EDLs at the anode and cathode, respectively; and $R_{EDL1}$ and $R_{EDL2}$ are the resistances of the EDLs at the anode and cathode, respectively; $R_{ION}$ is the bulk ion resistance; $CPE_{INT}$ and $R_{INT}$ represent the constant phase element and resistance owing to transverse internal effects.

Figure 6a and 6b show the resistance vs. frequency (Bode plot) and phase vs. frequency, respectively from EIS analysis from $10^{-2}$ to $10^5$ Hz. Additionally, the Nyquist plot is shown in Figure S5, and the real and imaginary components of impedance are plotted vs frequency in Figure S6. From the Bode plot in Figure 6a, compared to the pristine device without $LiPF_6$ addition, we observed that the impedance in the low frequency region is lower for 0.5% and 1% $LiPF_6$ and higher for 2% and 5%. The optimized $LiPF_6$ concentration (0.5%) produces the most conductive film at low frequencies. The corner frequencies reveal the changing timescales of ion redistribution in these devices. The pristine device corner frequency of 45 Hz is increased to 800 Hz with 0.5% LiPF$_6$, and decreased for higher concentrations. This implies that the 0.5% LiPF$_6$ concentration optimizes ionic conductivity in addition to overall conductivity.

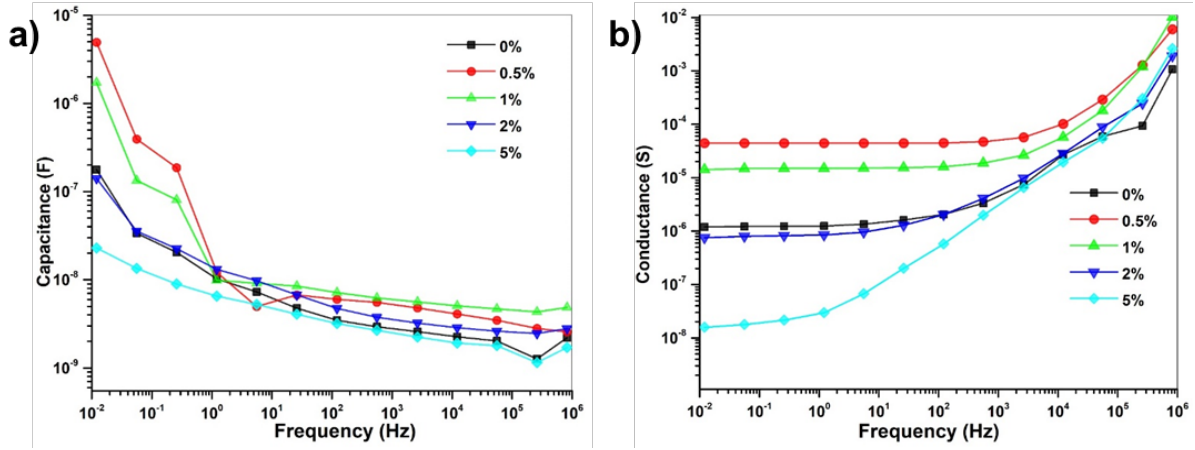

Figure 7. Capacitances and conductance extracted from EIS study. (a) Capacitance of PeLEC with various LiPF$_6$ concentrations vs frequency. (b) Conductance of PeLEC with various LiPF$_6$ concentrations vs frequency.

We calculated the capacitance and conductance of each PeLEC as a function of frequency using following formula:

$$\frac{1}{Z} = \frac{i_{ac}}{v_{ac}} = |Z|e^{i\phi} = G(\omega) + i\omega C(\omega) \qquad (4)$$

Where $Z$, $\omega$, $G$, and $C$ correspond to impedance, angular frequency, conductance and capacitance, respectively. Intuitively, due to the range of ionic conductivity effects imparted by LiPF$_6$, ionic transport strongly depends on the frequency of the applied AC electric field. To characterize this, capacitance vs frequency and conductance vs frequency of PeLEC devices with various LiPF$_6$ concentration are plotted in Figure 7a and 7b, respectively. From Figure 7a, at higher frequencies, the devices exhibited similar capacitances due to the lack of ionic motion in this region. However, at lower frequencies, we observed a significant rise in capacitance for all devices, indicating ionic migration and EDL formation. Accordingly, PeLEC devices with 0.5% LiPF$_6$ showed the highest increase in capacitance at low frequency. At lower frequencies, the conductance is enhanced in 0.5% and 1.0% LiPF$_6$ blends relative to the pristine device but is reduced at higher concentrations. Similar trends are also seen in impedance and capacitance vs voltage, as shown in Figure S7. These observations provided further evidence that an optimal LiPF$_6$ concentration maximizes ionic conductivity and ionic transport.

To further understand the impedance data, we used an equivalent circuit model (Figure 6c) previously used by Bastatas et al.[20] In this model, $R_E$ represents the electrical resistance of the active layer including charge injection resistance; $C_{GEO}$ represents the geometric capacitance of the overall device; two capacitors ($C_{EDLA}$ and $C_{EDLC}$) in parallel with two resistors ($R_{EDLA}$ and $R_{EDLC}$) represent the capacitances and resistances of the electric double layers at the anode (*EDLA*) and cathode (*EDLC*); $R_{ION}$ indicates the bulk ion resistance; $CPE_{INT}$ and $R_{INT}$ are the constant phase element and resistance that represent transverse internal effects such as grain boundaries in the polycrystalline active layer; $L_{cab}$ is the inductance of the external cables.

The parameters from fitting with the aforementioned equivalent circuit model and calculated physical quantities are shown in Table 1. The values of dielectric constants were calculated by approximating geometrical capacitance as a parallel plate capacitor. We observed an increase in dielectric constant with LiPF$_6$ concentration from 0 to 1 wt%, followed by a subsequent decrease with further increase of LiPF$_6$ concentration from 2 to 5 wt%. The variation in dielectric constant provides a quantitative description of the degree of ionic polarizability, in effect, ionic mobility.

**Table 1.** Parameters derived from equivalent circuit modeling with the model shown in Figure 6c.

| LiPF$_6$ [wt.%] | C$_{GEO}$ [nF] | Thickness [nm] | Dielectric Constant | C$_{EDLA}$ [nF] | C$_{EDLC}$ [nF] | W$_{EDLA}$ [nm] | W$_{EDLC}$ [nm] | Film Conductivity [Sm$^{-1}$] |
|---|---|---|---|---|---|---|---|---|
| 0% | 1.9 | 125 | 8.9 | 51 | 20.9 | 4.6 | 11.4 | 4.25 ×10$^{-6}$ |
| 0.5% | 3.4 | 130 | 16.6 | 88 | 75.4 | 4.99 | 5.80 | 1.06 ×10$^{-5}$ |
| 1% | 4.46 | 130 | 21.8 | 35 | 24.7 | 16.4 | 23.5 | 8.75 ×10$^{-6}$ |
| 2% | 2.53 | 125 | 11.9 | 8.91 | 9.19 | 20.1 | 34.3 | 4.87 ×10$^{-6}$ |
| 5% | 1.7 | 125 | 8 | 6.68 | 5.3 | 35.8 | 40.1 | 2.28 ×10$^{-6}$ |

In order to understand the influence of ionic additive on ionic conductivity, we used the extracted $R_{ION}$ values to obtain the ionic conductance $G_{ION}$, by fitting the aforementioned equivalent circuit. The ionic conductivity can be determined using the following equation:

$$\sigma = \frac{G_{ion}d}{A}, \qquad (2)$$

where $\sigma$ is film conductivity, $d$ is the thickness of the active layer and $A$ is the device area. We observed a significant increase in film conductivity with increasing LiPF$_6$ concentration from 0 to 0.5 wt%, followed by a modest increase from 1 to 2 wt%, and finally a decrease at 5 wt% (Table 1). Enhancement in ionic conductivity signifies the role of LiPF$_6$ salt as a source of mobile ions in PeLECs.

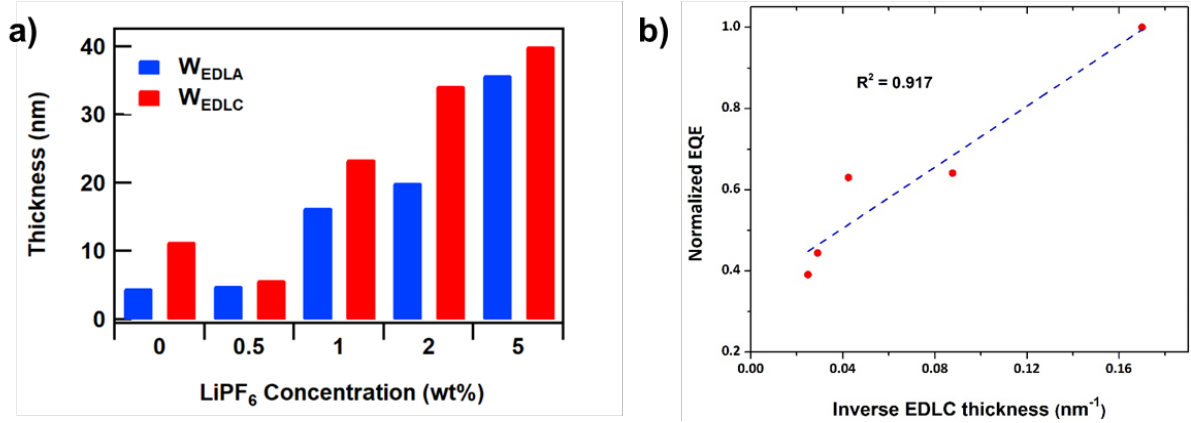

Figure 8. Double layer thicknesses extracted from EIS study and relation to EQE. (a) Anodic and cathodic EDL thickness extracted from EIS vs LiPF$_6$ concentration in PeLECs. (b) Normalized EQE vs inverse cathodic EDL thickness ($W_{EDLC}^{-1}$).

To explore the influence of the externally added mobile Li$^+$ and PF$_6^-$ ions on EDL formation, we calculated the EDL width ($d_{EDL}$) near each electrode by the following formula while assuming a uniform dielectric constant of the active layer:

$$d_{EDL} = \frac{d C_{GEO}}{C_{EDL}} \qquad (3)$$

where $C_{EDL}$ is the capacitance of the EDL, $d$ is thickness of the active layer and $C_{GEO}$ is the extracted geometric capacitance of the active layer from the equivalent circuit model. We observed a decrease in the width of the EDLs with increasing LiPF$_6$ concentration from 0 to 0.5 wt% followed by an increase with further increase of LiPF$_6$ concentration. Since PeLEC performance is strongly correlated with EDL formation, we show the EDL width of all fabricated PeLECs with various LiPF$_6$ concentrations in Figure 8a. Asymmetric and higher EDL widths are seen for the pristine device (0 wt% of LiPF$_6$) but symmetric and smaller EDL widths were observed in the device with 0.5 wt% LiPF$_6$. A smaller EDL width indicates higher charge accumulation for higher electric fields and lower injection barrier widths at the electrodes, and symmetry of the EDL widths promotes balanced electron and hole injection for efficient light emission. For higher LiPF$_6$ concentrations, the EDL widths become imbalanced and increased with increasing LiPF$_6$ concentration from 1-5 wt% - a range that is consistent with that determined experimentally as less optimal for device performance.

To emphasize the role of the EDL width and charge injection, we plotted maximum EQE obtained at a constant current driving current vs the inverse of cathodic EDL width (Figure 8b). A linear relationship is revealed, showing that a thinner EDL results in more efficient PeLEC operation. This correlation strongly substantiates our assertion of the key role of optimal concentration LiPF$_6$ addition in EDL formation for improved PeLEC device performance.

## 3. Conclusion

In this work, we have demonstrated that PeLECs utilizing an optimized fraction of LiPF$_6$ salt additive exhibit enhanced stability. At 0.5 wt% LiPF$_6$, devices exhibit 100 h operation in excess of 800 cd m$^{-2}$ under constant current driving, achieving a maximum luminance of 3260 cd m$^{-2}$ and power efficiency of 9.1 Lm W$^{-1}$. This performance extrapolates to a 6700 h luminance half-life from 100 cd m$^{-2}$, a 5.6-fold improvement over devices with no lithium salt additive. Analysis of LEC operation under constant voltage driving reveals three current regimes, with lithium addition strongly enhancing current in the second and third regimes. The third regime correlates degradation of luminance with decreased current. These losses are mitigated by LiPF$_6$ addition, an effect postulated to arise from preservation of perovskite structure. To further understand the influence of lithium additives, electrochemical impedance spectroscopy with equivalent circuit modeling is performed. Capacitance and conductance are strongly enhanced in the low frequency regions by lithium addition. Electrical double layer widths from ionic redistribution are minimized at 0.5 wt% LiPF$_6$ and inversely correlate with efficient performance. These results demonstrate that an optimize fraction of salt additive increases device stability and efficiency through improved double layer formation and retention of perovskite structure.

## 4. Experimental Section

**Materials:** Lead (II) bromide (PbBr$_2$; 99.99% trace metal basis), Cesium bromide (CsBr; 99.99%) and Polyethylene Oxide (PEO; M.W. > 5,000,000) were all purchased from Alfa Aesar. Lithium Hexafluorophosphate (LiPF$_6$; 99.99%) and Dimethyl Sulfoxide (DMSO; anhydrous > 99.9 % ) were purchased from Sigma Aldrich.

**Perovskite Solution Preparation:** The CsPbB$_3$ precursor solution was prepared by dissolving PbBr$_2$: CsBr (1:1.5 molar ratio) in DMSO and kept overnight for dissolution. PEO (10mg/ml) were prepared in DMSO solution. Dissolved CsPbBr$_3$ and PEO were mixed in 100:80 weight ratio. LiPF$_6$ salt (4mg/ml in DMSO) was added to this solution in various ratios. The final CsPbBr$_3$:PEO:LiPF$_6$ precursor solution was prepared with four different weight ratios (0.5%, 1%, 2% and 5%) of LiPF$_6$.

**Device Fabrication:** The overall PeLEC device architecture is as follows: ITO/PEDOT:PSS/ active layer/ LiF/Al. The active layer consisted of CsPbBr$_3$, PEO and different concentrations of LiPF$_6$. Prepatterned indium tin oxide (purchased from Thin Film Devices, Anaheim, CA) were cleaned in a sequence of non-ionic detergent wash, water bath sonication, and UV ozone treatment. Aqueous poly(3,4-ethylenedioxythiophene):polystyrene sulfonate (PEDOT:PSS) solutions (1.3−1.7%, Clevios AI 4083) were filtered through a 0.45 μm GHP filter and then spin-coated to obtain a ∼20 nm thick film on the ITO-coated glass substrates. These films were subsequently annealed at 100 °C for 10 minutes in a dry N$_2$ filled glovebox. The prepared active layer precursor solution was spin casted onto PEDOT: PSS at 1500rpm followed by vacuum treatment for 90 seconds and then thermally annealed at 70 °C for 5 minutes. The active layer thicknesses were generally 125-130 nm. To deposit the top electrode, samples were transferred to a vacuum chamber, and 10 Å LiF and 800 Å Al were deposited using a shadow mask that defined 12 devices per substrate, each with a 3 mm$^2$ device area.

**LEEC Device Testing:-**

The electrical and radiant flux characteristics were obtained with a custom multiplexer testing station capable of measuring 16 light emitting devices simultaneously. In brief, this instrument served as a current or voltage source and measuring unit and captured radiant flux with a calibrated Hamamatsu photodiode (S2387-1010R) for each device. The device slides were driven at a constant 3.5 V (99 mA compliance) for 12 h or at constant current. EIS measurements were obtained with a 760D electrochemical analyzer from CH Instruments (Austin, TX). For the EIS study, unbiased devices were zero-biased for 1 min and probed with alternating voltage (100 mV amplitude). Impedance spectra were then fitted using the Gamry Echem Analyst software provided by Gamry instruments. The fitting was accomplished with the equivalent circuit shown in Figure 5A, which is similar to Bastatas et al.[20] All parameters were allowed to vary but bound to overall device resistance, device thickness, the ranges of values typically reported in the literature. Normalized fitting errors were found to be vary from 0.0025 to 0.003. Critical parameter errors were generally between 0.1 and 30%.

**Supporting Information**
Supporting Information is available from the Wiley Online Library or from the author.

**Acknowledgements**
J.D.S. acknowledges support from the National Science Foundation (ECCS 1906505). Q.G. acknowledges support from the Welch Foundation (AT-1992-20190330). A.Z. acknowledges support from the Welch Foundation (AT-1617) and from the Ministry of Education and Science of the Russian Federation (14.Y26.31.0010). A.Z. also thanks financial support of Russian Federation grants No K2-2019-014 in NUST MISIS and 14.Y26.31.0010 in ITMO University.
Received: ((will be filled in by the editorial staff))
Revised: ((will be filled in by the editorial staff))
Published online: ((will be filled in by the editorial staff))

Perovskite light emitting electrochemical cells show 100 h operation in excess of 800 cd m$^{-2}$ and extrapolated lifetimes of 6700 h at 100 cd m$^{-2}$ with an optimal concentration of a lithium salt additive. Electrochemical impedance spectroscopy reveals lithium additives enhance efficiency through improved electrical double layer formation.

**Electroluminescence**

Aditya Mishra, Masoud Alahbakhshi, Ross Haroldson, Lyndon D. Bastatas, Qing Gu, Anvar A. Zakhidov and Jason D. Slinker*

**Enhanced Operational Stability of Perovskite Light-Emitting Electrochemical Cells Leveraging Ionic Additives**

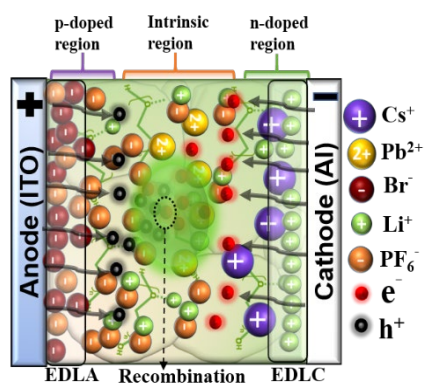

[1]     a) L. N. Quan, B. P. Rand, R. H. Friend, S. G. Mhaisalkar, T. W. Lee, E. H. Sargent, *Chem. Rev.* **2019**, *119*, 7444-7477; b) J. Z. Song, L. M. Xu, J. H. Li, J. Xue, Y. H. Dong, X. M. Li, H. B. Zeng, *Adv. Mater.* **2016**, *28*, 4861-4869; c) Y. H. Dong, Y. Gu, Y. S. Zou, J. Z. Song, L. M. Xu, J. H. Li, F. Xue, X. M. Li, H. B. Zeng, *Small* **2016**, *12*, 5622-5632.

[2]     a) X. F. Zhao, J. D. A. Ng, R. H. Friend, Z. K. Tan, *ACS Photonics* **2018**, *5*, 3866-3875; b) S. D. Stranks, R. L. Z. Hoye, D. Di, R. H. Friend, F. Deschler, *Adv. Mater.* **2019**, *31*, 1803336.

[3]     a) J. P. Wang, N. N. Wang, Y. Z. Jin, J. J. Si, Z. K. Tan, H. Du, L. Cheng, X. L. Dai, S. Bai, H. P. He, Z. Z. Ye, M. L. Lai, R. H. Friend, W. Huang, *Adv. Mater.* **2015**, *27*, 2311-2316; b) H. C. Cho, S. H. Jeong, M. H. Park, Y. H. Kim, C. Wolf, C. L. Lee, J. H. Heo, A. Sadhanala, N. Myoung, S. Yoo, S. H. Im, R. H. Friend, T. W. Lee, *Science* **2015**, *350*, 1222-1225; c) J. J. Si, Y. Liu, Z. F. He, H. Du, K. Du, D. Chen, J. Li, M. M. Xu, H. Tian, H. P. He, D. W. Di, C. Q. Ling, Y. C. Cheng, J. P. Wang, Y. Z. Jin, *ACS Nano* **2017**, *11*, 11100-11107; d) K. B. Lin, J. Xing, L. N. Quan, F. P. G. de Arquer, X. W. Gong, J. X. Lu, L. Q. Xie, W. J. Zhao, D. Zhang, C. Z. Yan, W. Q. Li, X. Y. Liu, Y. Lu, J. Kirman, E. H. Sargent, Q. H. Xiong, Z. H. Wei, *Nature* **2018**, *562*, 245-248; e) Y. C. Ling, Y. Tian, X. Wang, J. C. Wang, J. M. Knox, F. Perez-Orive, Y. J. Du, L. Tan, K. Hanson, B. W. Ma, H. W. Gao, *Adv. Mater.* **2016**, *28*, 8983-8989.

[4]     a) L. F. Zhao, K. M. Lee, K. Roh, S. U. Z. Khan, B. P. Rand, *Adv. Mater.* **2019**, *31*, 805836; b) C. Li, N. N. Wang, A. Guerrero, Y. Zhong, H. Long, Y. F. Miao, J. Bisquert, J. P. Wang, S. Huettner, *J. Phys. Chem. Lett.* **2019**, *10*, 6857-6864; c) Y. Zou, Z. Yuan, S. Bai, F. Gao, B. Sun, *Materials Today Nano* **2019**, *5*, 100028.

[5]     W. D. Xu, Q. Hu, S. Bai, C. X. Bao, Y. F. Miao, Z. C. Yuan, T. Borzda, A. J. Barker, E. Tyukalova, Z. J. Hu, M. Kawecki, H. Y. Wang, Z. B. Yan, X. J. Liu, X. B. Shi, K. Uvdal, M. Fahlman, W. J. Zhang, M. Duchamp, J. M. Liu, A. Petrozza, J. P. Wang, L. M. Liu, W. Huang, F. Gao, *Nat. Photonics* **2019**, *13*, 418-426.

[6]     a) L. N. Quan, Y. B. A. Zhao, F. P. G. de Arquer, R. Sabatini, G. Walters, O. Voznyy, R. Comin, Y. Y. Li, J. Z. Fan, H. R. Tan, J. Pan, M. J. Yuan, O. M. Bakr, Z. H. Lu, D. H. Kim, E. H. Sargent, *Nano Lett.* **2017**, *17*, 3701-3709; b) S. T. Zhang, C. Yi, N. N. Wang, Y. Sun, W. Zou, Y. Q. Wei, Y. Cao, Y. F. Miao, R. Z. Li, Y. Yin, N. Zhao, J. P. Wang, W. Huang, *Adv. Mater.* **2017**, *29*, 1606600; c) X. L. Yang, X. W. Zhang, J. X. Deng, Z. M. Chu, Q. Jiang, J. H. Meng, P. Y. Wang, L. Q. Zhang, Z. G. Yin, J. B. You, *Nat. Commun.* **2018**, *9*, 570; d) H. Tsai, W. Y. Nie, J. C. Blancon, C. C. Stoumpos, C. M. M. Soe, J. Yoo, J. Crochet, S. Tretiak, J. Even, A. Sadhanala, G. Azzellino, R. Brenes, P. M. Ajayan, V. Bulovic, S. D. Stranks, R. H. Friend, M. G. Kanatzidis, A. D. Mohite, *Adv. Mater.* **2018**, *30*, 1704217; e) T. Wu, J. Li, Y. Zou, H. Xu, K. Wen, S. Wan, S. Bai, T. Song, J. A. McLeod, S. Duhm, F. Gao, B. Sun, *Angew. Chem.-Int. Ed.*, DOI: 10.1002/anie.201914000; f) J. Q. Li, S. G. R. Bade, X. Shan, Z. B. Yu, *Adv. Mater.* **2015**, *27*, 5196-5202; g) L. Q. Zhang, X. L. Yang, Q. Jiang, P. Y. Wang, Z. G. Yin, X. W. Zhang, H. R. Tan, Y. Yang, M. Y. Wei, B. R. Sutherland, E. H. Sargent, J. B. You, *Nat. Commun.* **2017**, *8*, 15640; h) S. J. Lee, J. H. Park, B. R. Lee, E. D. Jung, J. C. Yu, D. Di Nuzzo, R. H. Friend, M. H. Song, *J. Phys. Chem. Lett.* **2017**, *8*, 1784-1792; i) Z. F. Shi, Y. Li, Y. T. Zhang, Y. S. Chen, X. J. Li, D. Wu, T. T. Xu, C. X. Shan, G. T. Du, *Nano Lett.* **2017**, *17*, 313-321; j) F. Yan, J. Xing, G. C. Xing, L. Quan, S. T. Tan, J. X. Zhao, R. Su, L. L. Zhang, S. Chen, Y. W. Zhao, A. Huan, E. H. Sargent, Q. H. Xiong, H. V. Demir, *Nano Lett.* **2018**, *18*, 3157-3164.

[7]     V. Prakasam, D. Tordera, H. J. Bolink, G. Gelinck, *Adv. Opt. Mater.* **2019**, *7*, 1900902.

[8]     a) Y. B. Yuan, J. S. Huang, *Accounts Chem. Res.* **2016**, *49*, 286-293; b) Z. G. Xiao, Y. B. Yuan, Y. C. Shao, Q. Wang, Q. F. Dong, C. Bi, P. Sharma, A. Gruverman, J. S. Huang, *Nat. Mater.* **2015**, *14*, 193-198; c) M. L. Lai, A. Obliger, D. Lu, C. S. Kley, C. G. Bischak, Q. Kong, T. Lei, L. T. Dou, N. S. Ginsberg, D. T. Limmer, P. D. Yang, *Proc. Natl. Acad. Sci. U. S. A.* **2018**, *115*, 11929-11934.

[9]     a) R. S. Sanchez, V. Gonzalez-Pedro, J. W. Lee, N. G. Park, Y. S. Kang, I. Mora-Sero, J. Bisquert, *J. Phys. Chem. Lett.* **2014**, *5*, 2357-2363; b) O. Almora, I. Zarazua, E. Mas-Marza, I. Mora-Sero, J. Bisquert, G. Garcia-Belmonte, *J. Phys. Chem. Lett.* **2015**, *6*, 1645-1652; c) B. Chen, M. J. Yang, S. Priya, K. Zhu, *J. Phys. Chem. Lett.* **2016**, *7*, 905-917; d) G. C. Xing, N. Mathews, S. S. Lim, N. Yantara, X. F. Liu, D. Sabba, M. Gratzel, S. Mhaisalkar, T. C. Sum, *Nat. Mater.* **2014**, *13*, 476-480; e) L. Protesescu, S. Yakunin, M. I. Bodnarchuk, F. Krieg, R. Caputo, C. H. Hendon, R. X. Yang, A. Walsh, M. V. Kovalenko, *Nano Lett.* **2015**, *15*, 3692-3696; f) Y. B. Yuan, T. Li, Q. Wang, J. Xing, A. Gruverman, J. S. Huang, *Sci. Adv.* **2017**, *3*, e1602164.

[10]    a) J. W. Lee, S. G. Kim, J. M. Yang, Y. Yang, N. G. Park, *APL Mater.* **2019**, *7*, 041111; b) M. H. Futscher, J. M. Lee, L. McGovern, L. A. Muscarella, T. Y. Wang, M. I. Haider, A. Fakharuddin, L. Schmidt-Mende, B. Ehrler, *Mater. Horizons* **2019**, *6*, 1497-1503; c) A. Walsh, D. O. Scanlon, S. Y. Chen, X. G. Gong, S. H. Wei, *Angew. Chem.-Int. Edit.* **2015**, *54*, 1791-1794; d) J. Haruyama, K. Sodeyama, L. Y. Han, Y. Tateyama, *J. Am. Chem. Soc.* **2015**, *137*, 10048-10051; e) N. Vicente, G. Garcia-Belmonte, *Adv. Energy Mater.* **2017**, *7*, 51700710; f) Y. F. Sun, M. Kotiuga, D. Lim, B. Narayanan, M. Cherukara, Z. Zhang, Y. Q. Dong, R. H. Kou, C. J. Sun, Q. Y. Lu, I. Waluyo, A. Hunt, H. Tanaka, A. N. Hattori, S. Gamage, Y. Abate, V. G. Pol, H. Zhou, S. Sankaranarayanan, B. Yildiz, K. M. Rabe, S. Ramanathan, *Proc. Natl. Acad. Sci. U. S. A.* **2018**, *115*, 9672-9677; g) S. Miyoshi, T. Akbay, T. Kurihara, T. Fukuda, A. T. Staykov, S. Ida, T. Ishihara, *J. Phys. Chem. C* **2016**, *120*, 22887-22894; h) C. Aranda, A. Guerrero, J. Bisquert, *ACS Energy Lett.* **2019**, *4*, 741-746.

[11]    M. F. Ayguler, M. D. Weber, B. M. D. Puscher, D. D. Medina, P. Docampo, R. D. Costa, *J. Phys. Chem. C* **2015**, *119*, 12047-12054.

[12]    H. M. Zhang, H. Lin, C. J. Liang, H. Liu, J. J. Liang, Y. Zhao, W. G. Zhang, M. J. Sun, W. K. Xiao, H. Li, S. Polizzi, D. Li, F. J. Zhang, Z. Q. He, W. C. H. Choy, *Adv. Funct. Mater.* **2015**, *25*, 7226-7232.

[13]    B. M. D. Puscher, M. F. Ayguler, P. Docampo, R. D. Costa, *Adv. Energy Mater.* **2017**, *7*, 1602283.

[14]    P. Andricevic, X. Mettan, M. Kollar, B. Nafradi, A. Sienkiewicz, T. Garma, L. Rossi, L. Forro, E. Horvath, *ACS Photonics* **2019**, *6*, 967-975.

[15]    M. Alahbakhshi, A. Mishra, R. Haroldson, A. Ishteev, J. Moon, Q. Gu, J. D. Slinker, A. A. Zakhidov, *ACS Energy Lett.* **2019**, 2922-2928.

[16]    a) J. D. Slinker, J. A. DeFranco, M. J. Jaquith, W. R. Silveira, Y.-W. Zhong, J. M. Moran-Mirabal, H. G. Craighead, H. D. Abruna, J. A. Marohn, G. G. Malliaras, *Nat. Mater.* **2007**, *6*, 894-899; b) J. C. de Mello, *Phys. Rev. B* **2002**, *66*, 235210.

[17]    a) Y. Shen, D. D. Kuddes, C. A. Naquin, T. W. Hesterberg, C. Kusmierz, B. J. Holliday, J. D. Slinker, *Appl. Phys. Lett.* **2013**, *102*, 203305; b) L. D. Bastatas, K. Y. Lin, M. D. Moore, K. J. Suhr, M. H. Bowler, Y. L. Shen, B. J. Holliday, J. D. Slinker, *Langmuir* **2016**, *32*, 9468-9474; c) K. Y. Lin, L. D. Bastatas, K. J. Suhr, M. D. Moore, B. J. Holliday, M. Minary-Jolandan, J. D. Slinker, *Acs Applied Materials & Interfaces* **2016**, *8*, 16776-16782.

[18]    a) P. Wellmann, M. Hofmann, O. Zeika, A. Werner, J. Birnstock, R. Meerheim, G. F. He, K. Walzer, M. Pfeiffer, K. Leo, *J. Soc. Inf. Disp.* **2005**, *13*, 393-397; b) X. L. Dai,

Z. X. Zhang, Y. Z. Jin, Y. Niu, H. J. Cao, X. Y. Liang, L. W. Chen, J. P. Wang, X. G. Peng, *Nature* **2014**, *515*, 96-99.
[19] L. Zhao, J. Gao, Y. L. Lin, Y. W. Yeh, K. M. Lee, N. Yao, Y. L. Loo, B. P. Rand, *Adv. Mater.* **2017**, *29*, 1605317.
[20] L. D. Bastatas, M. D. Moore, J. D. Slinker, *ChemPlusChem* **2018**, *83*, 266-273.

Supporting Information

**Enhanced Operational Stability and Efficiency of Perovskite Light Emitting Electrochemical Cells Using Ionic Additives**

*Aditya Mishra, Masoud Alahbakhshi, Ross Haroldson, Lyndon D. Bastatas, Qing Gu, Anvar A. Zakhidov and Jason D. Slinker\**

**Table S1** Comparative performances of the best-in-class perovskite-based LEDs and LECs.

| Device structure | Single (S) or multilayer (M) | Light Emitting Mechanism | Turn on voltage (V) | Device Stability with operation condition | Ref* |
|---|---|---|---|---|---|
| ITO / Pero-PEO-composite /In (Ga,Au) | S | LED | 3 | 1hr (L30) at constant 2.7 V | [1] |
| ITO/ZnO/PVP/perovskite/CBP/MoOx/Al | M | LED | ~ 3 | 0.01 h (L50) at constant 5V driving | [2] |
| ITO/PEDOT: PSS/perovskite/SPW111/LiF/Ag | M | LED | 2.4 @ 0.1 cd/m$^2$ | 3.9 h (L70) | [3] |
| ITO/PolyTPD/FAPbBr$_3$/TpBi/LiF/Al | M | LED | 4 | 0.1 h (EQE50) at constant 10 mA/cm$^2$ | [4] |
| Au/p-MgNiO/CsPbBr$_3$/PMMA/n-MgZnO/n$^+$-GaN | M | LED | ~ 4 | 10 h (L80) | [5] |
| ITO/PEDOT:PSS/MAPbBr$_3$/B$_3$PYMPM:TpBi/B3PYMPM: Cs$_2$CO$_3$/Al | M | LED | ~ 3 | 0.1 h (L50) | [6] |
| ITO/PEDOT:PSS/CsPbBr$_3$:MABr/B$_3$PYMPM/LiF/Al | M | LED | ~ 2.7 | 100 h @100 cd/m$^2$ (L50) with constant 5mA | [7] |
| ITO/NiO$_X$+PVK+TFB/ CsPbBr$_3$+LiBr/TpBi/LiF/Al | M | LED | ~2.4 | 108 h @100 cd/m$^2$ (L50) | [8] |
| **Our Work** | **S** | **LEC** | **1.95** | **6700 h @100 cd/m$^2$ (L50) with constant 1.5mA** | |

*References are listed at the end of this file.

**Table S2** Summarized ionic parameters relevant to PeLECs from literature.

| Ion type and ionic process | Activation Energy (eV) | | Diffusion coefficient ($cm^2 s^{-1}$) | Mobile ions concentration ($cm^{-3}$) | Characteristic Times | Ref[*] |
|---|---|---|---|---|---|---|
| | Theory | Experimental | | | | |
| $MA^+$, $FA^+$ | 0.46-1.12 | Depends on device fabrication | $(3.4\pm3.3)\ 10^{-12}$ | $(1.3\pm0.8)\ 10^{16}$ | $\tau=12\pm0.4$ min (EDL by drift of $MA^+$) | [9] |
| $Cs^+$ (EDL by cations) | 0.8-2.31 | | $(3.1\pm2.8)\ 10^{-9}$ | $(1.1\pm0.9)\ 10^{15}$ | | |
| $I^-$, $Br^-$ (EDL by halides) | 0.08-0.58 | $0.29\pm0.06$ | | | $\tau=30\pm1.5$ s (EDL by halide drift) | [10] |
| $Li^+$ | 0.24 | | $10^{-7}$ | Varies | - | [11,12] |
| $PF_6^-$ | 0.366 | | $1.18 \times 10^{-13}$ | Varies | - | [13] |
| P-doped layer formation | | | | | $\tau=1.6\pm0.1$ h (growth of charge carriers doped layer via ions) | [14] |
| n-doped layer formation | | | | | | |

*References are listed at the end of this file.

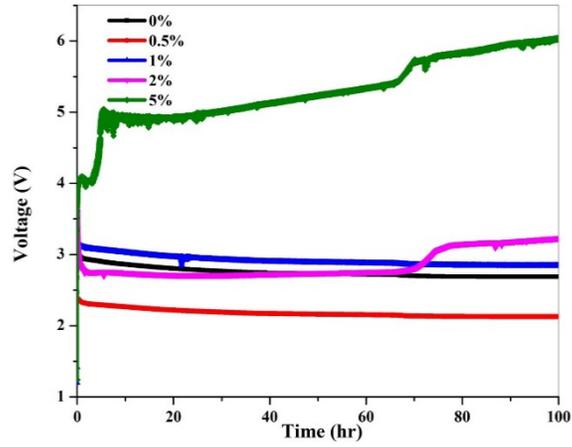

Figure S1 Voltage vs time of PeLEC devices with various LiPF$_6$ concentrations under constant current driving (0.050 A/cm$^2$).

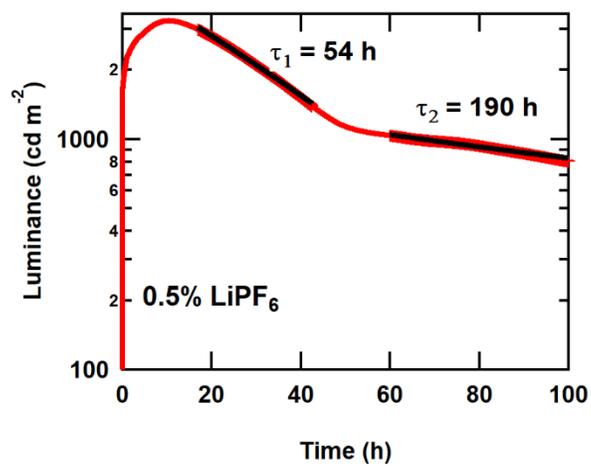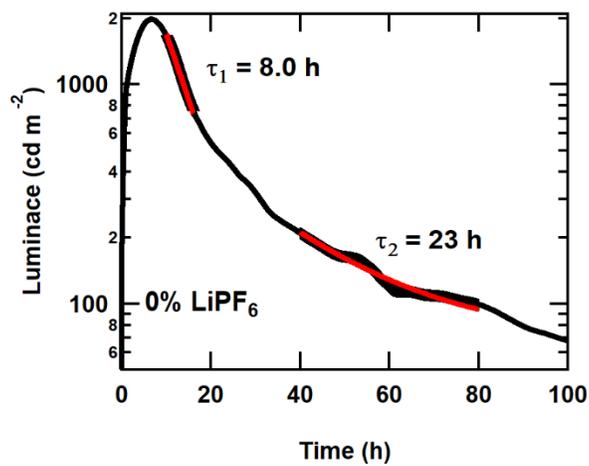

Figure S2 Luminance versus time for PeLEC devices with 0.5% and 0% LiPF6 concentrations. Exponential fits are made in the regions indicated by thicker data points and contrasted fit lines.

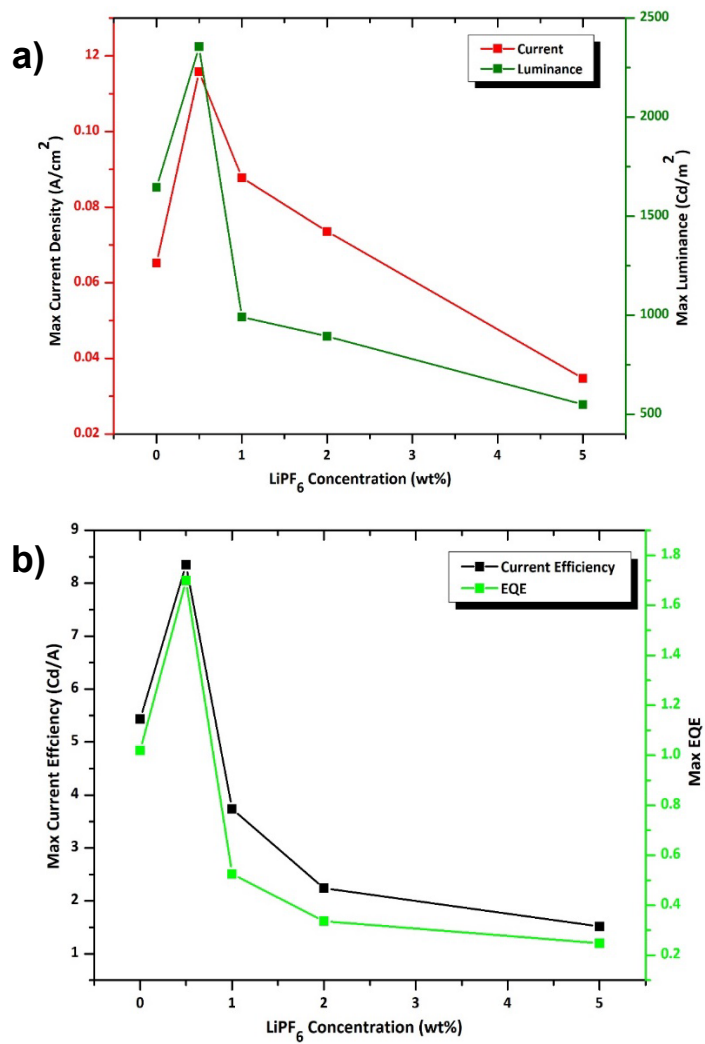

Figure S3 Performance of PeLEC devices with various $LiPF_6$ concentrations under constant voltage driving (3.5V) (a) Current density (left axis) and luminance (right axis) versus $LiPF_6$ concentration (wt%) in PeLEC. (b) Maximum current efficiency and maximum external quantum efficiency obtained during constant voltage driving (3.5V) versus $LiPF_6$ concentration (wt%) in PeLEC.

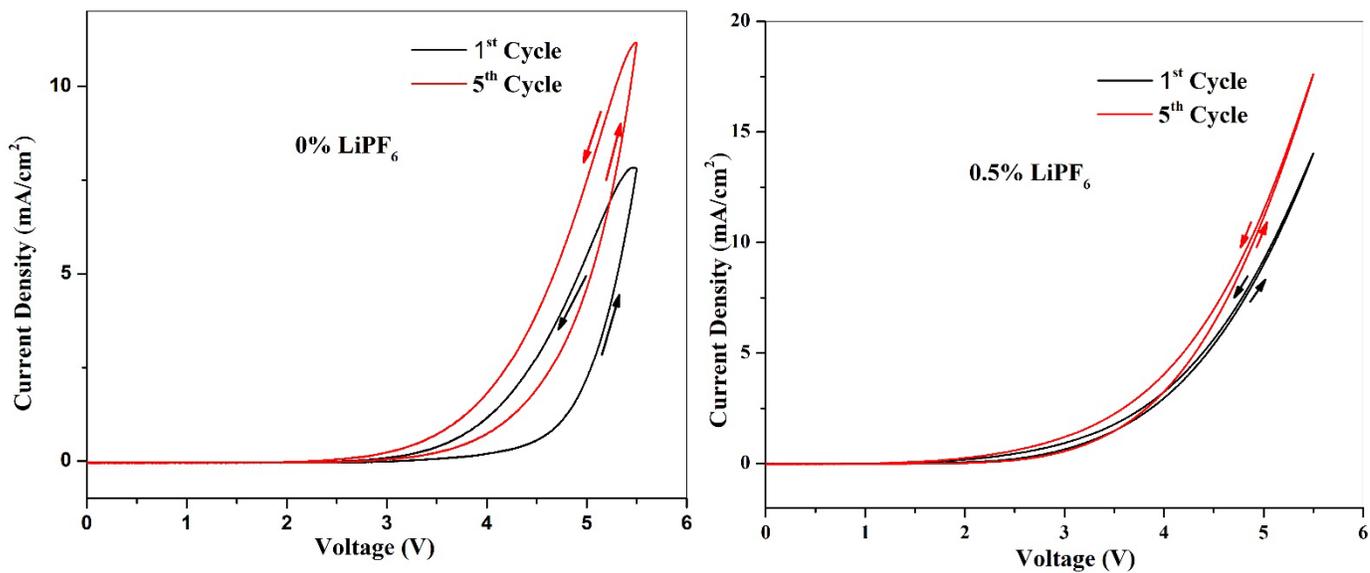

Figure S4. Current density (J) versus voltage (V) of PeLEC devices with 0% and 0.5% of LiPF$_6$. The cyclic J-V curve clearly shows that in the optimized ratio of Li (0.5%), the hysteresis state decreases significantly.

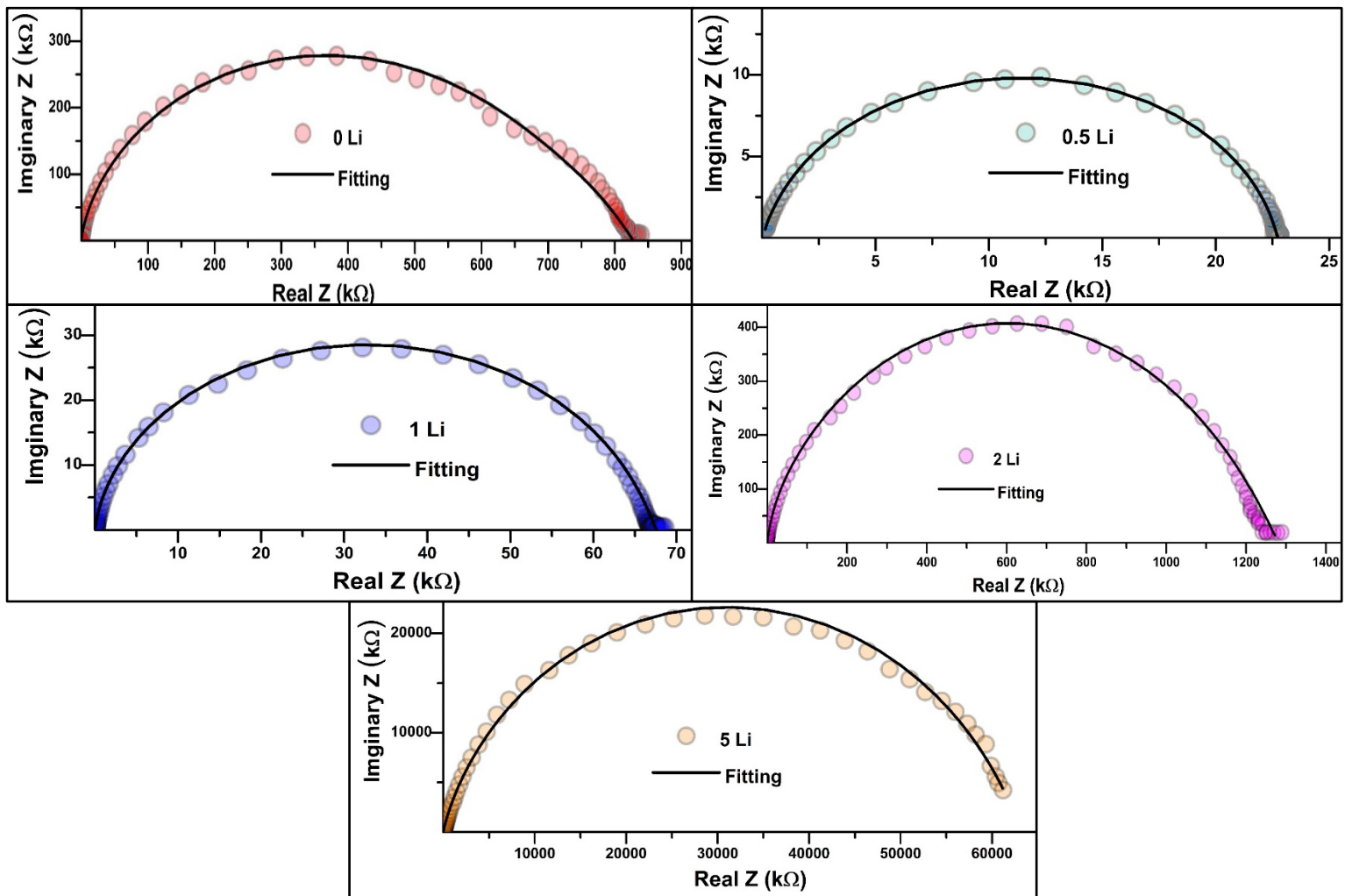

Figure S5 Nyquist plots of imaginary versus real components of impedance for PeLEC devices with various LiPF$_6$ concentrations. Solid lines represent fitting according to the equivalent circuit model discussed in the main text.

**Table S1** Parameters extracted from the impedance spectra of devices with different concentration of $LiPF_6$ concentration.

| Sample (%$LiPF_6$) | $C_{GEO}$ (nF) | $R_E$ (MΩ) | $R_{ion}$ (KΩ) | $R_{EDLA}$ (MΩ) | $R_{EDLC}$ (MΩ) | $R_{INT}$ (MΩ) | $Q_{INT}$ (C×10$^{-9}$) | $α_{INT}$ |
|---|---|---|---|---|---|---|---|---|
| 0% | 1.90 | 4.62 | 8.50 | 9.39 | 15.9 | 0.729 | 8.67 | 0.757 |
| 0.5 | 3.40 | 0.0309 | 4.08 | 106 | 91.3 | 0.0801 | 14.2 | 0.793 |
| 1 | 4.46 | 0.120 | 5.05 | 67.7 | 56.9 | 0.150 | 24.6 | 0.707 |
| 2 | 2.53 | 5.62 | 8.56 | 0.542 | 0.0489 | 0.160 | 37.3 | 0.731 |
| 5 | 1.70 | 95.3 | 18.3 | 0.00244 | 0.000128 | 18.8 | 9.11 | 0.749 |

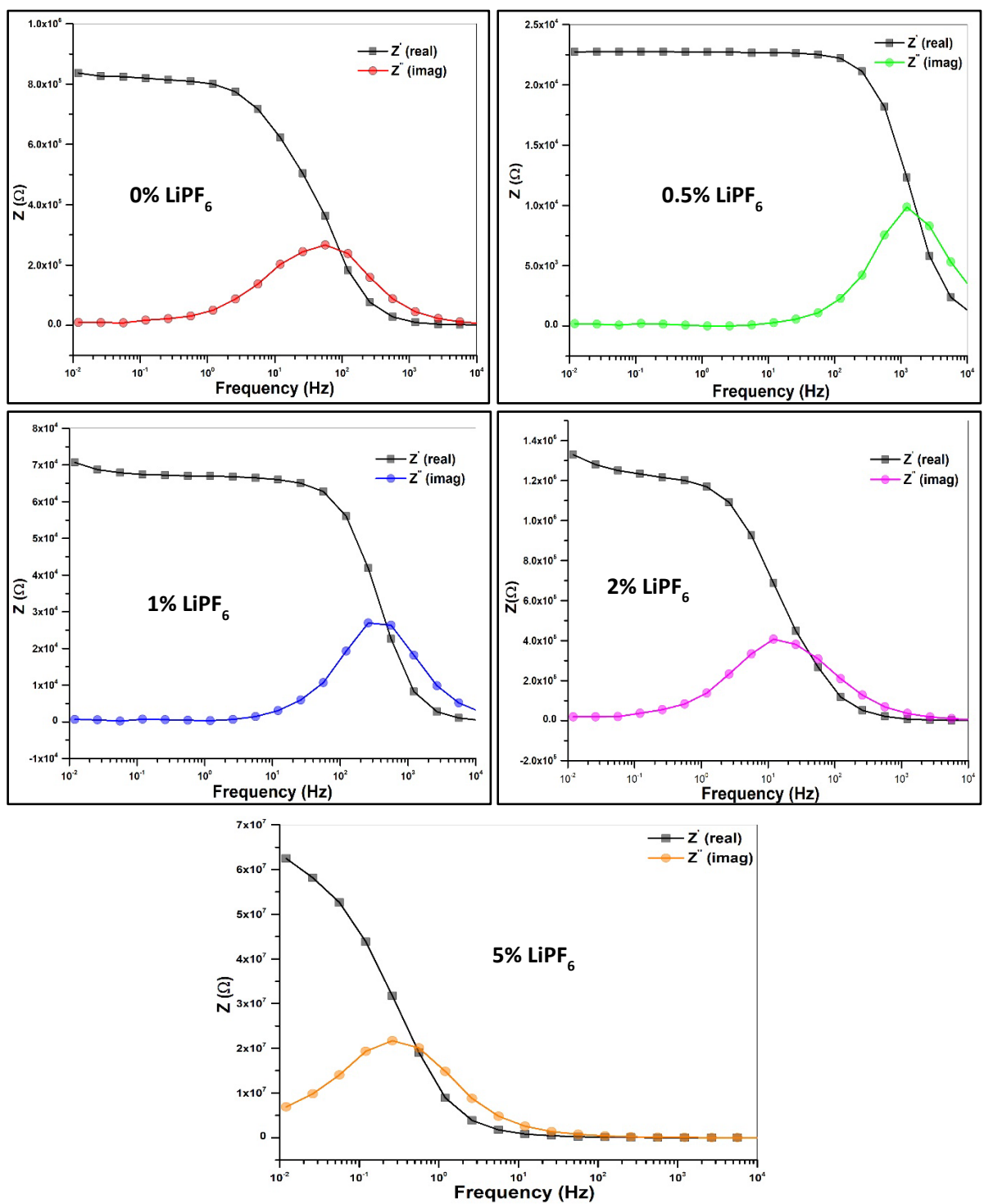

Figure S6 Real and imaginary components of impedance vs frequency plots of PeLEC devices with various LiPF$_6$ concentration.

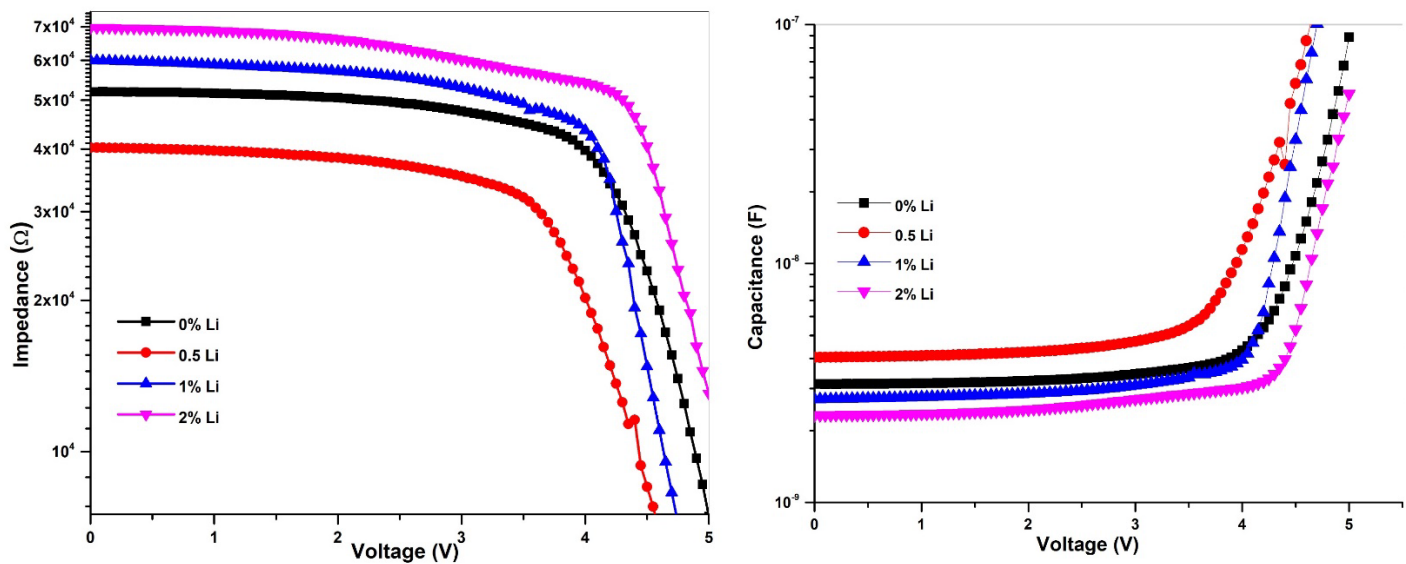

Figure S7 (a) Impedance versus voltage (V) of PeLEC devices with various $LiPF_6$ concentrations. (b) Capacitance versus voltage (V) of PeLEC devices with various $LiPF_6$ concentrations.

**References**

[1] J. Q. Li, S. G. R. Bade, X. Shan, Z. B. Yu, *Adv. Mater.* **2015**, *27*, 5196-5202.
[2] L. Q. Zhang, X. L. Yang, Q. Jiang, P. Y. Wang, Z. G. Yin, X. W. Zhang, H. R. Tan, Y. Yang, M. Y. Wei, B. R. Sutherland, E. H. Sargent, J. B. You, *Nat. Commun.* **2017**, *8*, 15640.
[3] S. J. Lee, J. H. Park, B. R. Lee, E. D. Jung, J. C. Yu, D. Di Nuzzo, R. H. Friend, M. H. Song, *J. Phys. Chem. Lett.* **2017**, *8*, 1784-1792.
[4] L. F. Zhao, K. M. Lee, K. Roh, S. U. Z. Khan, B. P. Rand, *Adv. Mater.* **2019**, *31*, 805836.
[5] Z. F. Shi, Y. Li, Y. T. Zhang, Y. S. Chen, X. J. Li, D. Wu, T. T. Xu, C. X. Shan, G. T. Du, *Nano Lett.* **2017**, *17*, 313-321.
[6] F. Yan, J. Xing, G. C. Xing, L. Quan, S. T. Tan, J. X. Zhao, R. Su, L. L. Zhang, S. Chen, Y. W. Zhao, A. Huan, E. H. Sargent, Q. H. Xiong, H. V. Demir, *Nano Lett.* **2018**, *18*, 3157-3164.
[7] K. B. Lin, J. Xing, L. N. Quan, F. P. G. de Arquer, X. W. Gong, J. X. Lu, L. Q. Xie, W. J. Zhao, D. Zhang, C. Z. Yan, W. Q. Li, X. Y. Liu, Y. Lu, J. Kirman, E. H. Sargent, Q. H. Xiong, Z. H. Wei, *Nature* **2018**, *562*, 245-248.
[8] T. Wu, J. Li, Y. Zou, H. Xu, K. Wen, S. Wan, S. Bai, T. Song, J. A. McLeod, S. Duhm, F. Gao, B. Sun, *Angew. Chem.-Int. Ed.*, DOI: 10.1002/anie.201914000.
[9] M. H. Futscher, J. M. Lee, L. McGovern, L. A. Muscarella, T. Y. Wang, M. I. Haider, A. Fakharuddin, L. Schmidt-Mende, B. Ehrler, *Mater. Horizons* **2019**, *6*, 1497-1503.
[10] B. M. D. Puscher, M. F. Ayguler, P. Docampo, R. D. Costa, *Adv. Energy Mater.* **2017**, *7*, 1602283.
[11] N. Vicente, G. Garcia-Belmonte, *Adv. Energy Mater.* **2017**, *7*, 51700710.
[12] Y. F. Sun, M. Kotiuga, D. Lim, B. Narayanan, M. Cherukara, Z. Zhang, Y. Q. Dong, R. H. Kou, C. J. Sun, Q. Y. Lu, I. Waluyo, A. Hunt, H. Tanaka, A. N. Hattori, S. Gamage, Y. Abate, V. G. Pol, H. Zhou, S. Sankaranarayanan, B. Yildiz, K. M. Rabe, S. Ramanathan, *Proc. Natl. Acad. Sci. U. S. A.* **2018**, *115*, 9672-9677.
[13] S. Miyoshi, T. Akbay, T. Kurihara, T. Fukuda, A. T. Staykov, S. Ida, T. Ishihara, *J. Phys. Chem. C* **2016**, *120*, 22887-22894.
[14] C. Aranda, A. Guerrero, J. Bisquert, *ACS Energy Lett.* **2019**, *4*, 741-746.